\documentclass{ws-procs975x65}

\begin{document}

\title{Definition and Classification of Singularities in GR: Classical and Quantum}

\author{D. A. KONKOWSKI}

\address{Department of Mathematics, \\
U.S. Naval Academy, \\ 
Annapolis, Maryland, 21402, USA\\ 
E-mail: dak@usna.edu}

\author{T. M. HELLIWELL}

\address{Department of Physics, \\ 
Harvey Mudd College, \\
Claremont, California, 91711, USA\\
E-mail: T\_Helliwell@HMC.edu}

\maketitle

\abstracts{
We will briefly review the definition and classification of classical and quantum singularities 
in general relativity. Examples of classically singular spacetimes that do not have quantum 
singularities will be given. We will present results on quantum singularities in quasiregular
spacetimes. We will also show that a strong repulsive "potential" near the classical singularity
can turn a classically singular spacetime into a quantum mechanically nonsingular spacetime.}

\section{Introduction}
After defining classical and quantum singularities examples of each will be given and their relationship discussed. This conference proceeding is based on [\refcite{KH}] and  [\refcite{HK}].

\section{Classical Singularities}
A classical singularity is indicated by incomplete geodesics or incomplete paths of bounded acceleration [\refcite{HE}] in a maximal spacetime. Since, by definition, a spacetime is smooth, all irregular points (singularities) have been excised; a singular point is a boundary point of the spacetime.

Classical singularities have been classified by Ellis and Schmidt [\refcite{ES}] into three basic types: quasiregular, non-scalar curvature, and scalar curvature. The mildest is quasiregular and the strongest is scalar curvature. A singular point $q$ is a quasiregular singularity if all components of the Riemann tensor $R_{abcd}$ evaluated in an orthonormal frame parallel propogated along an incomplete geodessic ending at $q$ are $C^{0}$ (or $C^{0-}$). In other words, the Riemann tensor components tend to finite limits (or are bounded). On the other hand, a singular point $q$ is a curvature singularity if some component is not bounded in this way. If all scalars in $g_{ab}$, the antisymmetric tensor $\eta_{abcd}$ and $R_{abcd}$ nevertheless tend to a finite limit (or are bounded), the singularity is non-scalar, but if any scalar is unbounded, the point $q$ is a scalar curvature singularity.

\section{Quantum Singularities}
Horowitz and Marolf [\refcite{HM}] have defined a static spacetime as quantum mechanically singular if the spatial portion of the Klein-Gordon wave operator is not essentially self-adjoint on a $C_{0}^{\infty}$ domain in $L^{2}$, a Hilbert space of square integrable functions. In this case the evolution of the test quantum wave packet is not uniquely determined by the initial wavefunction. Using this definition they found that Reissner-Nordstr\"{o}m, negative mass Schwarzschild, and the 2D cone remain singular when probed by quantum wave packets, but certain orbifold spacetimes, extreme Kaluza-Klein black holes, and some other string theory examples are nonsingular. In another context, Kay and Studer [\refcite{KS}] have also shown that neither the 2D cone nor a thin idealized cosmic string have essentially self-adjoint wave operators.

\section{Quasiregular Spacetimes}
A quasiregular spacetime is a spacetime with a classical quasiregular singularity. Conical singularities such as those in the 2D cone and the 4D idealized cosmic string spacetime are quasiregular singularities. Thus some of the mildest true singularities have been shown to be singular quantum mechanically as well as classically. What about other static quasiregular spacetimes? We examined [\refcite{HK}] quasiregular spacetimes with dislocations and disclinations  [\refcite{PS}] including the exceedingly unusual Galtsov/Letelier/Tod spacetime [\refcite{GL}, \refcite{Tod}]. Broad classes of spacetimes with both screw dislocations and disclinations were found to be quantum mechanically singular - the wave modes ``felt" the singularity (Horowitz and Marolf [\refcite{HM}] had noted this effect in spherically symmetric spacetimes).

\section{Cylindrical Spacetimes}
To study this effect further we considered a class of static, cylindrical metrics with timelike singularities [\refcite{KH}],
$$ds^2 = -dt^2 +dr^2 + r^{2a}d \phi ^2 + r^{2b}dz^2,$$
\noindent where $r$ is a radial coordinate and $\phi$ is an angular coordinate with the usual ranges. These metrics have classical scalar curvature singularities unless $a=1$ and $b=0$ (flat spacetime in cylindrical polar coordinates). What about quantum singularities? To determine this we used the test equation $(A\pm i)\Psi=0$ and studied its solutions. The operator is essentially self-adjoint if only one solution is square integrable near the classical singularity at $r=0$. Separating variable we considered the radial part of equation in Schrodinger form,
$$F'' + \left[ \pm i - \frac{(\frac{a+b}{2})[(\frac{a+b}{2}) -1]}{r^2} - \frac{m^2}{r^{2a}} - 
\frac{k^2}{r^{2b}}\right]F=0.$$
\\
Assume $m=0$, $k=0$, for simplicity. Near $r=0$, if $a+b<2$, the ``potential'' is attractive, while if $a+b \ge 2$, the ``potential'' is repulsive. Near $r=0$, one solution of the original equation goes like a constant (and is thus $L^2$ using the appropriate measure) and the other goes like $r^{1-(a+b)}$ (and is thus $L^2$ if $a+b<3$). We therefore see that these cylindrical spacetimes are 
quantum mechanically singular if $a+b<3$ (except for Minkowski spacetime) and quantum mechanically nonsingular if $a+b \ge 3$ (or Minkowski spacetime). These cylindrical spacetimes are thus quantum mechanically nonsingular if the spacetime metric induces a very repulsive potential.

\section{Conclusions}
Quantum singularities have been shown to appear in broad classes of spacetimes with quasiregular singularities and scalar curvature singularities. However, classically singular spacetimes with strong repulsive barriers shielding the singularity may be quantum mechanically nonsingular [\refcite{HM}, \refcite{HK}, \refcite{KH}]. In these spacetimes quantum wave packets simply bounce off the barrier and never reach the singularity. Geodesics are the geometric optics limit of infinite frequency waves and only in that limit is the singularity reached.

\section*{Acknowledgements}
One of us (DAK) was partially funded by NSF grants PHY-9988607 and PHY-0241384 to
the U.S. Naval Academy. She also thanks Queen Mary, University of London,
where some of this work was carried out.

\end{document}